\documentclass[10pt,draft]{dis03}
\usepackage{epsf,amsmath}

\begin{document}

\title{DIS 2003 - An Overview\thanks{Summary talk at DIS 2003, St. Petersburg, 23-27 April 2003.}}

\author{A. H. Mueller\footnote{This research is supported in part by the U.S. Department of Energy.} \\
Department of Physics \\
Columbia University\\
New York, New 10027
}

\maketitle

\begin{abstract}
\noindent An overview of DIS 2003 is given with a focus on small-x physics and the interrelationship between deep inelastic scattering and heavy ion physics which furnishes much of the basic information for understanding high-density QCD matter.  In addition to small-x physics, topics covered  include single spin asymmetries, light-cone gauge choices, strange quarks in the proton and non-global observables in jet decays.
\end{abstract}

\section{Why do (study) deep inelastic scattering?}

Deep inelastic scattering is the sharpest and most precise probe of the QCD partonic structure of hadrons. Non small-x inelastic and elastic electron scattering are probably the best ways to probe the nature of the dominant part of the proton's wavefunction, and such measurements connect with fundamental nonperturbative QCD via lattice gauge theory.

Small-x DIS probes a new region of QCD, the regime of partonic saturation or the Color Glass Condensate, where the gauge potential $A_\mu$ becomes as large as $1/g_{QCD}.$  We recall the principal domains of QCD probed in DIS are:\\
\\
\indent (i)  $\alpha$ small, $A_\mu\simeq {\sqrt{\alpha}} \leftrightarrow$ perturbation theory.\\

\indent (ii)  $\alpha = 0(1), A_\mu =O(1) \leftrightarrow$ traditional nonperturbative QCD\\

\indent (iii)  $\alpha$ small, $A\simeq 1/{\sqrt{\alpha}} \leftrightarrow$ new domain of high-density QCD (Color Glass Condensate/Saturation)\\
\\
In small-x QCD deep inelastic scattering and hadronic scattering are complementary and go well together.  While DIS is the better probe of the small-x dense partonic wavefunction of a proton or nucleus the matter in the wavefunction is not quite real.  Hadronic scattering, in particular high-energy heavy ion collisions, free dense partonic matter so that it becomes genuinely real, but probes here are very difficult to interpret precisely. What makes  DIS and heavy ion collisions fit so well together is the close correspondence between high-density (virtual) matter in small-x hadronic wavefunctions and high-density produced matter in a high-energy hadronic or heavy ion collision.  Of course in non small-x physics this close correspondence between the quarks in the wavefunction of a proton and the jets produced in deep inelastic scattering is the reason why the quark parton model is such a satisfying description of the structure of the proton.

Small-x deep inelastic scattering touches on one of the deepest parts of QCD, that of  high-field strengths and of high quantum occupancy of states, which has become one of he central areas of theoretical study.  As such deep inelastic scattering experiments, along with high-energy heavy ion collisions, are furnishing the crucial experimental input for achieving a more complete and deeper understanding of dense matter.

\section{Partons (a review)}
Since the idea of partons will play such an important role in the discussion which follows it is perhaps useful to take a few 
moments to review how they appear in the description of the $F_2$ structure  function in deep inelastic scattering. \\

Partons are manifest in deep inelastic scattering in light-cone gauge and in the Bjorken frame.  In the Bjorken frame the proton's momentum, P, and the virtual photon's momentum, $q,$ take the form

\begin{equation}
P_\mu=(P_0, P_1, P_2, P_3) \simeq (p + {m^2\over 2p}, 0, 0, p)
\end{equation}

\begin{equation}
q_\mu = (q_0, \underline{q}, 0)
\end{equation}

\noindent where  $p$  is taken very large and $q_0={P\cdot q\over p}$ becomes very small.  The scattering amplitude is illustrated in Fig.1 with the invariants  $x$ and $Q^2$ given as $Q^2 = - q_\mu q_\mu \simeq \underline{q}^2$ and $x = {Q^2\over 2P\cdot q} \simeq k/p.$

\vskip 10pt

\begin{figure}[!thb]
\begin{center}
\centerline{\epsfbox[0 0 139 137]{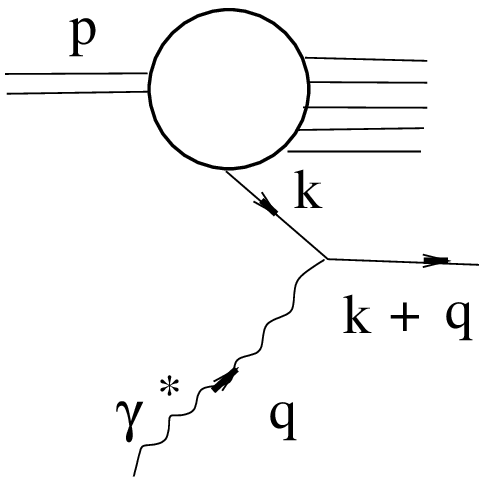}}
\caption{}
\end{center}
\end{figure}

In the Bjorken frame and in light-cone gauge the virtual photon is absorbed by the quark in a transverse spatial region

\begin{equation}
\Delta x_\perp \simeq 1/q_\perp = 1/Q
\end{equation}

\noindent and over a time

\begin{equation}
\Delta t =\left[E_k + E_q-E_{k + q}\right]^{-1} \simeq 1/Q
\end{equation}

\noindent so that the virtual photon makes a local (in transverse coordinate space) and instantaneous measurement of the charge carrying partons (quarks) in the proton.  In formulas

\begin{equation}
F_2(x, Q^2) = \sum_f e_f^2\left[ xq_f(x,Q^2) +x\bar{q}_f(x,Q^2)\right]
\end{equation}

\noindent with $e_f$ being the charge of a quark of flavor  $f.$  $q_f$ and $\bar{q}_f$ are the quark and antiquark number densities, for measurements made on a scale $\Delta x_\perp \sim 1/Q,$ in the proton.

\section{Saturation - Color Glass Condensate[1]}

Later I shall come back to a more detailed discussion of saturation, but it is perhaps useful to state early what the densities of quarks and gluons are in the Color Glass Condensate.  The essential idea is that in a light-cone wavefunction quark and gluon densities reach universal limits below an energy dependent transverse momentum scale. I think it is most illuminating to express these densities in terms of quantum occupation numbers.  Then for quarks\cite{Ler}

\begin{equation}
f_q \simeq {(2\pi)^3\over 2\cdot 2\cdot N_c} \ {dx[q + \bar{q}]\over d^2b_\perp d^2k_\perp} \simeq 1/\pi
\end{equation}

\noindent for $k_\perp^2 \leq \bar{Q}_s^2(x,b_\perp).$  In (6) $b_\perp$ is the impact parameter at which the quark densities are given.  So long as $k_\perp$ is greater than the inverse size of the hadron there should be no conflict in using both $b_\perp$ and $k_\perp$ to describe the quark. In (6) the two factors of 2 in the denominator correspond to the number of  spins and the fact that the density $[q + \bar{q}]$ involves both quarks and antiquarks.  $f_q$ is supposed to be a three-dimensional occupation number.  The longitudinal variables do not apppear explicitly because 

\begin{equation}
d\ell_z db_z \simeq d\ell_z/\ell_z\simeq dx/x
\end{equation}  

\noindent and the $dx$  is implicitly included in the definition of $[q + \bar{q}].$  Of course the fact that there is only one layer of quarks in the longitudinal direction and (7) makes use of the uncertainty principle means that (6) has an approximate character.  Neverthless, it is remarkable that quark phase space densities reach such a simple, dynamics independent, limit as given by (6), a limit which apparently is not related to the Pauli Principle.

For gluons the situation is a bit more complicated, and more interesting.  Here one expects[2-4]

\begin{equation}
f_g \simeq {(2\pi)^3\over 2(N_c^2-1)}\ {dxG\over d^2b_\perp d^2k_\perp}\ \simeq\ {c_1\over \alpha_sN_c} [\ell n Q_s^2/k_\perp^2 + c_2]
\end{equation}                                                                                 

\noindent for $k_\perp^2 \leq Q_s^2, $  and where the constants $c_1$ and $c_2$ are not yet understood. The gluon saturation momentum, $Q_s(x,b_\perp),$ and the quark saturation momentum, $\bar{Q}_s,$ are related by

\begin{equation}
\bar{Q}_s^2 = {C_F\over N_c}\ Q_s^2.
\end{equation}

\noindent In the most interesting region where most of the saturated gluons are located, $k_\perp \approx Q_s, f_g \sim {1\over \alpha N_c}$ corresponding to $A_\mu \sim {1\over {\sqrt{\alpha}}}.$  For protons in the HERA regime and for heavy nuclei in the RHIC energy regime $Q_s$ is estimated to be on the order of a $GeV.$

\section{Wavefunctions vs reactions}
Hadronic wavefunctions are not physical objects in a gauge theory such as QCD. Cross-sections, structure functions and distributions of particles in a reaction are the physical quantities in high-energy reactions.  It often happens that there is a close relationship between the wavefunction of a hadron and a measurable quantity if the wavefunction is calculated in an appropriate gauge in QCD. Thus the left-hand side of (5) is a measureable quantity, the $F_2$ structure function of the proton, while the right-hand side of (5) represents the density of quarks in a Fock space description of the proton's wavefunction in light-cone gauge.  It is often the case that it is technically easier to determine properties of a wavefunction than to determine properties of a high-energy reaction, though in some cases the opposite is true.  For that reason it is important to understand when, and to what extent, the properties of the wavefunction of a nucleon or nucleus lead to predictions for high-energy reactions.  In the following subsections we are going to explore in some detail the relationship between wavefunctions and reactions. We begin with the example of the quark sea in a large nucleus.

\subsection{Quark sea in a large nucleus} 

A good example in which to see the intricate and subtle relationship which can exist between the wavefunction of a  hadron and a high-energy reaction is  the quark sea in a large nucleus.  Begin by considering the light-cone wavefunction of a large nucleus in an approximation where the gluons in the wavefunction are treated semiclassically[5,6] and where only a single quark-antiquark pair is explicitly considered.  The light-cone perturbation theory graphs contributing to this pair are shown in Fig.2[6] where, in addition, we attach at the end  a virtual photon which turns the wavefunction into a sea contribution to $F_2.$  (For the moment ignore the dashed line ending in indices $\alpha$ and $\beta$.)  It is this picture to which (6) applies and says that we expect a limit to the phase space density of sea quarks which is independent of the size of the nucleus. One can, in principle, measure the transerse momentum dependence of the sea quark distribution of the nucleus, and this is made most manifest if one takes $q$ to have no transverse momentum. However,  calculating the graphs in Fig.2 is very difficult and  it is much easier to change gauges and view the process as the scattering of a high-energy virtual photon on a nucleus at rest with the calculation of the transverse momentum dependence of the current quark jet being done in, say, a covariant gauge.

\vskip 10pt

\begin{figure}[!thb]
\begin{center}
\epsfbox[0 0 205 98]{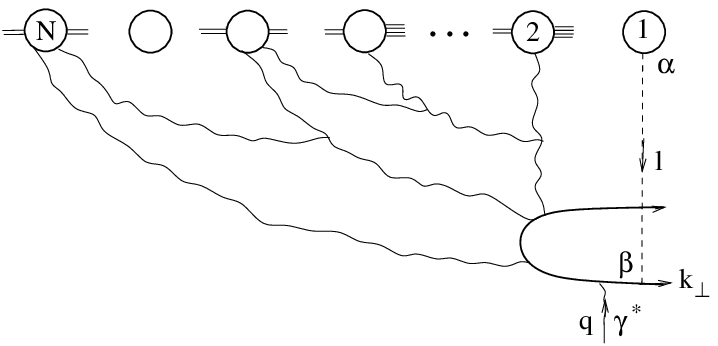}
\caption{}
\end{center}
\end{figure}

In a covariant gauge calculation of quark production the process looks very different.  It now appears that the virtual photon first breaks up into a quark-antiquark pair, the sea quark-antiquark pair in the nuclear wavefunction description, which then carries out scattering through one and  two gluon exchanges with the nucleons in the nucleus.  This is now a straightforward Glauber-type calculation whose graphs are illustrated in Fig.3.  While an analytic form for the distribution in $k_\perp$ is easy to write down we quote here only the integrated, over $k_\perp,$ result which is proportional to the sea contribution to the $F_2$ structure function, the dominant contribution at small values of $ x.$   The result is[2,7]
\vskip 10pt

\begin{figure}[!thb]
\begin{center}
\epsfbox[0 0 221 85]{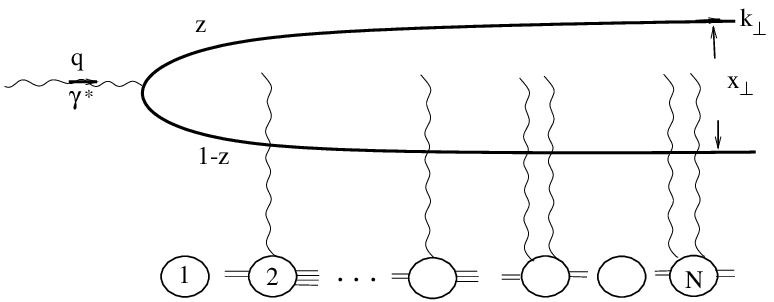}
\caption{}
\end{center}
\end{figure}

\begin{equation}
F_2={Q^2\over 8\pi^2\alpha_{em}} \int d^2x_\perp \int_0^1 dz \sum_f e_f^2\vert\psi_T(x_\perp,z,Q)\vert^2d^2b_\perp 2(1-S(b_\perp, x_\perp, x))
\end{equation}

\noindent where

\begin{equation}
S=e^{-\sigma_{dp}T(b_\perp)/2}
\end{equation}

\noindent with

\begin{equation}
T(b_\perp) = 2{\sqrt{R^2-b_\perp^2}}\ \rho
\end{equation}

\noindent and[8]

\begin{equation}
\sigma_{dp} = {\pi^2\alpha\over N_c}\ x_\perp^2 xG(x, 4/x_\perp^2)
\end{equation}

\noindent and where $\rho$ is the nuclear density and $\sigma_{dp}$ is the cross-section for a dipole of size $x_\perp$ to scatter on a proton.  $\psi_T$ is the lowest order quark-antiquark wavefunction of the virtual photon.

From (10) and (11) it is clear that $F_2$ cannot grow indefinitely as the nuclear size increases but will reach a limit, independent on the details of the QCD dynamics, when\ $S$ \ becomes zero.  The unitarity limit expressed by the eikonal form of \ $S$ \ is what causes the limiting distribution for the quark distribution (6) when the reaction is viewed in terms of the wavefunction pictured in Fig.2.  Thus parton saturation is equivalent to the unitarity limit for dipole nucleus scattering.  In the present case saturation is very hard to see directly from the light-cone perturbation theory graphs in Fig.2 while the unitarity limit for scattering is easy to obtain.  The direct connection between the wavefunction shown in Fig.2 and the reaction producing a quark-antiquark pair allows one here to deduce a general property of the light-cone wavefunction.

Finally, Eq.(10) is very close to the Golec-Biernat W\"usthoff[9] model which we shall discuss a bit later.  To arrive at the Golec-Biernat W\"ustoff model simply make the replacements

\begin{equation}
d^2b_\perp \to \sigma_0.
\end{equation}

\noindent and

\begin{equation}
S \rightarrow e^{-x_\perp^2\bar{Q}_s^2/4}.
\end{equation}

\noindent Indeed comparing (15) with (11) and using (12) and (13) leads to

\begin{equation}
Q_s^2 \equiv {N_c\over C_F} \bar{Q}_s^2 = {4\pi^2\alpha N_c\over N_c^2-1} 2{\sqrt{R^2-b_\perp^2}}\ \rho xG(x,Q_s^2)
\end{equation}

\noindent the form of the saturation momentum for a large nucleus in the semiclassical approximation, the McLerran-Venugopalan model.

\subsection{Light-cone gauges}

Consider again the process described in the previous section, but now with the final quark jet transverse momentum measured to be $k_\perp.$  In Fig.3 we have included all graphs, within our simple Glauber-type model.  However, in Fig.2, our light-cone gauge picture, one might worry that final state gluon exchanges, like that of the dashed line shown in that figure, might be important.  If such final state interactions are important then the simple picture relating the reaction, the cross-section for producing a jet having transverse momentum $k_\perp,$ to the light-cone wavefunction will be lost.  The following three statements briefly summarize the situation:

(i)  In a covariant gauge calculation exchanges like that of the dashed line in Fig.2 are important and the parton picture is not manifest.  The parton picture is not manifest because there is no simple relationship between observable reactions and the covariant gauge wavefunction of a hadron.

(ii)  If $k_\perp$ is integrated, that is for the structure function $F_2,$ final state interactions will cancel in any light-cone gauge[10,11].  For example,  the gluon exchange given by the dashed line in Fig.2, along with a similar final state gluon exchange in the complex conjugate amplitude, will cancel with two-gluon exchange final state interactions in the amplitude and in the complex conjugate amplitude.

(iii)  If $k_\perp$ is fixed, then final state interactions like the dashed line in Fig.2 will contribute in most light-cone gauges, but there is a particular light-cone gauge, the Kovchegov gauge[6], where final state interactions are absent[12].  The Kovchegov gauge has a propagator for the dashed line in Fig.2 given by

\begin{equation}
D_{\alpha\beta}(\ell) = {-i\over \ell^2+i\epsilon} [g_{\alpha\beta}-{\eta_\alpha\ell_\beta\over \ell_+-i\epsilon}\ - \ {\ell_\alpha\eta_\beta\over \ell_+ + i\epsilon}]
\end{equation}

\noindent with $\eta\cdot v = v_+$ for any  four-vector $v_\mu.$  The argument for the absence of final state interactions in the Kochegov gauge can be given in simple terms. In any light-cone gauge elastic rescattering corrections, like the dashed line in Fig.2, occur at a slow rate since the coupling of the  exchanged gluon at one end of the exchange must be to the transverse part of the current (the $\beta -$ vertex in Fig.2) which is small.  Thus elastic rescatterings occur over long periods of time.  The choice of $i\epsilon$'s in the light-cone gauge denominators allow one to put these rescattering corrections completely in the initial state (the Kovchegov gauge) or completely in the final state or in a mixture of the two.  The Kovchegov gauge puts all corrections into the initial state, that is in the wavefunction of the nucleus, thus preserving the simple relationship between the wavefunction and reactions where even certain details of the final state are determined.

\subsection{Single transverse spin asymmetries}

There has been much activity and new insight concerning single spin\newline
 asymmetries[13-15] in the past year or so.  This is a rich field with data[16-18] and challenges in deep inelastic scattering as well as in hadronic interactions.  Much of the discussion at the conference has been concerned with the level at which universality is present and the level at which the asymmetry survives when the scale of the hard scattering changes.

In order to illustrate some of the issues consider jet production in deep inelastic scattering as illustrated in Fig.4 where the proton $(P)$ is polarized.  In the proton rest system the single spin asymmetry corresponds to an expectation of the time reversal odd quantity

\begin{figure}
\begin{center}
\centerline{\epsfbox[0 0 168 95]{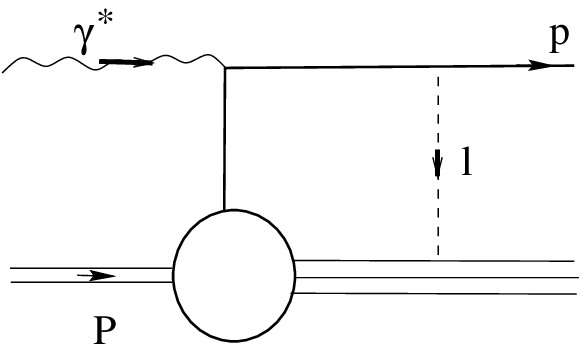}}
\caption{}
\end{center}
\end{figure}

\begin{equation}
i\left({\vec{\sigma}_P\over 2}\right) \cdot \left(\vec{q}{\rm  x} \vec{p}\right).
\end{equation}

\noindent (Of course one more easily measures an outgoing hadron of momentum $\vec{p},$ however, for purposes of illustration it is more convenient to consider $\vec{p}$ as an outgoing jet.)

In a generic light-cone or covariant gauge calculation the interaction of the outgoing jet with the fragments of the proton through gluon exchange (the $\ell$-line in Fig.4) is necessary in order to create a phase which simulates time reversal violation[14,15].  In the present case the $\ell$-interactions will not cancel in a generic light-cone gauge calculation because the transverse momentum of the jet is fixed and final state interactions affect detailed properties of the final state, although not overall rates.  However, in the Kovchegov gauge the $\ell$-line is absent and there are no final state interactions.  What are usually viewed as final state interactions are encoded in the wavefunction of the proton in the Kovchegov gauge.

Thus for single spin asymmetries, as measured in deep inelastic scattering, knowledege of the proton's wavefunction is sufficient to  predict the whole reaction probability.  This looks like factorization, and in a restricted sense it is.  However, it is not a complete QCD factorization, because the same factors do not occur in spin asymmetries measured in lepton-proton reactions and in proton-proton reactions[19].  To understand why this is the case consider the reaction proton (1) + proton (2) $\to \mu^+\mu^-(q)$ + anything.  The Kovchegov gauge has strong causality violating initial state reactions so that proton (1) can affect proton (2) before the production of the $\mu$-pair occurs.  Thus spin asymmetries measured in $\mu$-pair production will not be given in terms of a product of two Kovchegov- gauge proton wavefunctions and complete factorization is lost.

\subsection{Evolving into dense QCD}

There are two separate but equivalent procedures for dealing with the nonlinear properties of QCD.  A procedure for calculating the small-x wavefunction, including nonlinear high-density effects, has been developed by Jalilian-Marian, Iancu, McLerran, Weigert, Leonidov and Kovner (JIMWLK)[1,20-22] while an approach based on generalizing the BFKL equation to include nonlinear terms in scattering amplitudes has been developed by Balitsky and Kovchegov(BK)[23,24].  The Kovchegov equation is a simplified version of the Balitsky equations which are equivalent to the JIMWLK approach.

The idea of the JIMWLK approach to calculating small-x wavefunctions is relatively simple.  In a high-energy hadron or nucleus soft gluons are emitted from harder quarks and gluons as semiclassical (Weizs\"acker-Williams) emissions.  Then they add to the color charge fluctuations of the hadron furnishing new sources for emitting even softer gluons.  In the regime where the color charge fluctuations have not grown too large this is the light-cone wavefunction picture of BFKL evolution.  The JIMWLK procedure is general enough to apply also  when color charge fluctuations are very large.  After evolving through a region of rapidity\   $Y$\   one ends up with a description of the color charge fluctuations of a high-energy hadron as a function of rapidity, \ $y$,\ with $0 \leq y \leq Y.$  This information is expressed in terms of a probability distribution $W_Y[\rho]$ for finding a ``covariant gauge'' charge distribution $\rho(\underline{x}, x_-)$ in the hadron.  The equations are even simpler in terms of $\alpha(\underline{x}, x_-)$ where

\begin{equation}
\nabla_{\underline{x}}^2\alpha(\underline{x}, x_-) = - \rho(\underline{x}, x_-).
\end{equation}

\noindent $\alpha$ corresponds to a covariant gauge potential created by the sources in the high-energy hadron.

If the high-energy hadron has a large value of $p_+,$ that is it is travelling in the positive $z-$direction, then the JIMWLK procedure is very efficient for evaluating expectations of observables constructed from Wilson lines along the $x_-$ direction.  If O is an observable built up of Wilson lines of the type

\begin{equation}
V[\alpha(\underline{x})] = P e^{-ig\int_{-\infty}^\infty dx_-\alpha(\underline{x},x_-)},
\end{equation}

\noindent then

\begin{equation}
<O(\alpha) >_Y = \int D[\alpha]W_Y[\alpha]O(\alpha)
\end{equation}

\noindent with the basic (JIMWLK) equation

\begin{equation}
{d\over dY} W_Y[\alpha] = {\alpha_s\over 2} \int d^2x \int d^2y {\delta\over \delta\alpha^a(x)}\left[\eta_{\underline{x}\underline{y}}^{ab}{\delta W_Y\over \delta\alpha^b(y)}\right],
\end{equation}

\noindent where $\eta$ is given in terms of Wilson lines as a functional of $\alpha.$

Eq.(22) leads to the gluon saturation formula given in (8).  It also leads to an interesting property concerning the charge density-charge density correlator[25,26]

\begin{equation}
<\rho^a(k_\perp, b_\perp)\rho^b(-k_\perp, b_\perp)> = \delta_{ab} k_\perp^2/\pi,
\end{equation}

\noindent when $k_\perp^2/Q_s^2 \leq 1.$  Eq.(23) suggests that in the region where gluon densities are large there is an effective color charge shielding.  It is still a little mysterious as to how this comes about.  One expects color shielding in a wavefunction to be different than in a thermalized medium since there should not be any analog of the Debye mass in a wavefunction.  It is important, however,  to have good color charge shielding in a heavy ion wavefunction if one expects to have a smooth transition between the small-x gluons in the wavefunction and the gluons that ultlimately equilibrate after a heavy ion collision.

The other approach to small-x physics in the nonlinear regime is that of Balitsky and Kovchegov.  The procedure described by Balitsky for calculating high-energy scattering is equivalent to that of JIMWLK, but it deals with scattering amplitudes rather than wavefunctions.  The equation derived by Kovchegov is an approximation to the Balitsky equation.  It has the advantage that it is quite simple and elegant.  Let me sketch a derivation.

Consider the scattering of a dipople on a hadron at high-energies.  The dipole is eqivalent to two Wilson lines which form a color singlet before and after the scattering.  Knowledge of dipole hadron scattering is sufficient to determine the $F_1$ and $F_2$ structure functions of deep inelastic scattering.  Suppose the dipole consists of a quark part at $\underline{x}_1$ and an antiquark part at $\underline{x}_2$ with \ $Y$\ being the rapidity of the scattering.  Then $S(\underline{x}_1-\underline{x}_2, Y)$ is the elastic scattering $S-$matrix.  In the JIMWLK picture the evaluataion of $S$ is done by taking the expectation of the two Wilson lines making up the dipole in the field of the high-energy hadron as given by (21).  Again in the JIMWLK  picture the evolution in $Y$ is given by (22) where the wavefunction of the high-energy hadron changes with rapidity. However, one may equally change  $Y$ by a small amount, $dY,$ by fixing the high-energy hadron and instead by varying the rapidity of the dipole.  Changes in  $S$  then come about because the dipole may split into two dipoles before the scattering takes place in which case it is a two-dipole state which scatters on the high-energy hadron.  Probability conservation then requires a diminution of the single dipole state probability.  The process is pictured in Fig.5. The first term on the right-hand side of Fig.5 is a new object, the $S$-matrix for a two-dipole state to scatter on the high-energy hadron.  In order to get the Kovchegov equation one assumes that the two dipoles scatter independently on the hadron, in which case the equation becomes

\vskip 10pt 
\begin{figure}[!thb]
\begin{center}
\epsfbox[0 0 246 40]{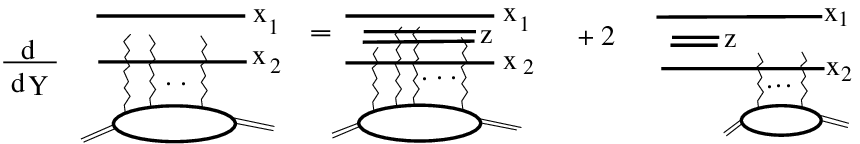}
\caption{}
\end{center}
\end{figure}

\begin{displaymath}
{dS(\underline{x}_1-\underline{x}_2, Y)\over dY}\ =\ {\alpha N_c\over 2\pi^2}\int {d^2z(\underline{x}_1-\underline{x}_2)^2\over (\underline{x}_1-\underline{z})^2(\underline{z}-\underline{x}_2)^2}
\end{displaymath}
\begin{equation}
\cdot \left[S(\underline{x}_1-\underline{z}, Y) S(\underline{z}-\underline{x}_2,Y) - S(\underline{x}_1-\underline{x}_2,Y)\right]
\end{equation}

\noindent in the large $N_c$ limit.  Eq.(24) is the Kovchegov equation[24].  It is probably the best simple equation governing the approach to the unitarity limit (saturation) at high-energy.  When  $S \simeq 1$ Eq.(24) is equivalent to the BFKL equation.  When  $S$ is small, that is deep in the saturation regime, (24) gives the correct form of the $S-$ matrix for dipole-dipole scattering but it misses the constant in the exponent by a factor of 2[27].  The Kovchegov equation is a good equation to study the transition from $S\simeq 1$ to $S \simeq 0$ but it cannot be expected to give exact results.

All in all I think there is now a pretty good theoretical understanding of saturation and unitarity limits in hard QCD.  The details of the transition between weak and strong scattering are not yet under control and, of course, there is an urgent need to confront our understanding with experimental data.

\subsection {Heavy Ion Collisions}

It is perhaps natural that deep inelastic scattering structure functions be closely related to the light-cone wavefunction of a hadron. It is less clear that there should be a relationship between light-cone wavefunctions and the early stages after a heavy ion collision.  On the other hand, such a relationship should not be too surprising.  The small-x light cone wavefunction is dominated by gluons which can be, roughly, categorized as belonging to two parts.  Firstly, there are those gluons having $k_\perp\leq Q_s$ and which form part of the Color Glass Condensate and, secondly, there are those gluons having $k_\perp > Q_s$ which are relatively dilute short time fluctuations in the wavefunction.  In a zero impact parameter collision of two identical heavy ions the gluons having $k_\perp > Q_s$ are unlikely to be freed since they are dilute amongst themselves during the collision and softer gluons cannot effectively free harder gluons.  On the other hand, one expects a good portion of the gluons having $k_\perp < Q_s$ to be freed since such a gluon in, say, the left-moving nucleus encounters a strong field, at a corresponding transverse wavelength, from the right-moving nucleus[28,29].  In addition, there may be gluons produced during the collision which were not part of either wavefunction, but such gluons will certainly not dominate those already in the wavefunctions of the colliding ions.

A picture of a heavy ion collision is shown in Fig.6 where one of the ions is taken at rest.  Just after the collision one can say that, roughly, those mid-rapidity gluons in the saturation regime of the fast ion are freed during the collision and go on to interact among themselves as they evolve toward equilibrium.  The high occupation numbers of these freed mid-rapidity gluons, $f_g \sim 1/\alpha,$ should make classical Yang-Mills dynamics adequate both for doing a calculation of just how many gluons are freed and for following the gluons in the early stages of their evolution toward equilibrium.  So far analytic calculations of the density of freed gluons have not been achieved, however, Krasnitz, Nara and Venugopalan and Lappi have done  numerical simulations of classical Yang-Mills dynamics starting from light-cone wavefunctions given by the McLerran-Venugopalan model[1,30-32].  Among other results they find that about one-half of the gluons in the wavefunction, see Fig.6, are freed in the collision.  This seems to justify our general picture and points to a smooth transition between the saturated gluons in the light-cone wavefunction and those gluons forming the matter just after a high-energy heavy ion collision.

\begin{figure}
\begin{center}
\epsfbox[0 0 263 155]{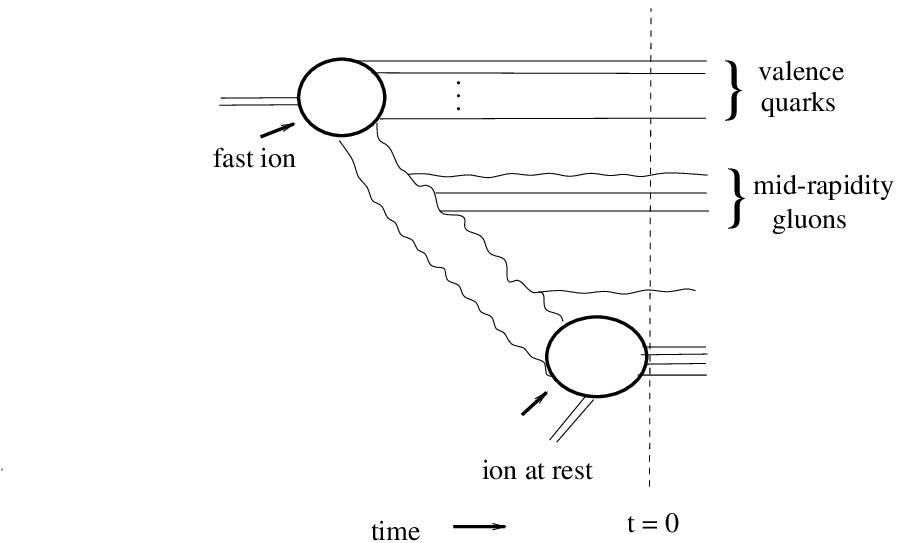}
\caption{}
\end{center}
\end{figure}

\section{Structure of the nucleon[33]}

Non small-x electron-nucleon scattering is a good tool to study the structure of the proton.  In addition to  traditional elastic and inelastic cross-sections a new element, deeply virtual compton scattering (DVCS), has recently received a lot of attention in this area.  Let me begin with the more traditional approaches.

\subsection{Traditional measures}

The spin-independent valence distributions  $u_{val}(x,Q^2)$ and $d_{val}(x,Q^2)$ as well as the spin-dependent distributions  $\Delta u_{val}$ and $\Delta d_{val}$ are clearly related to the structure and binding of the proton.  The role of the sea quarks, and in particular the strange sea, is not so clear. For example $s(x,Q^2)$ is measured and is not small.  Indeed,\  $s$\  is about of half the size of $u_{sea}$ or $d_{sea}.$  However, if $s\bar{s}$ pairs are just short time fluctuations in the proton they will not be important for the binding energy (the mass) of the proton and will play a very small role in building a realistic proton wavefunction.  On the other hand, a non zero $\Delta q_{sea}$ requires significant interaction of the sea with the rest of the proton and is thus a good measure of the importance of the sea for the static properties of the proton.  This is why there has been such intense interest in determining and interpreting the value of $\Delta s.$

Recently, two new measurements of the strange, $s$ minus $\bar{s},$ content of the elastic form factor of the proton have been done.  Again, these strange quark measurements can give non-zero results only if the strange sea has significant interaction with the rest of the proton.  A small value of the strange quark part of the elastic form factor would suggest that the strange quark content of the proton consists mostly of short time, benign, fluctuations while a large value would suggest that strange quarks play an essential role in the determination of the static properties of the proton.  The HAPPEX[34,37] experiment at Jefferson Lab. and the SAMPLE[35-37] experiment at Bates have evidence that in fact the strangeness content if the elastic form factor is quite small.  This is an interesting and important result and we should have much more detailed experimental evidence on this issue from a series of upcoming J-lab experiments.

\subsection{Deeply virtual compton scattering (DVCS)[38-40]}

In DVCS experiments, illustrated in Fig.7, one measures the elastic cross-section, and in some cases the actual elastic amplitude, for $\gamma^* +$ proton $\rightarrow \gamma +$ proton. The initial photon should be hard enough to be sure that single quarks in the proton are being measured.

\begin{figure}[!thb]
\begin{center}
\centerline{\epsfbox[0 0 120 90]{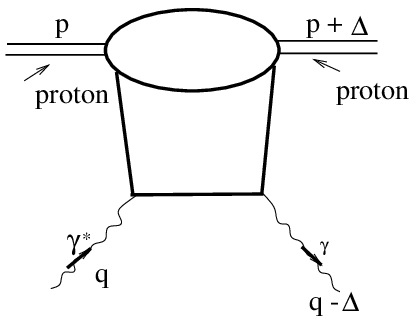}}
\caption{}
\end{center}
\end{figure}

There are potentially some extremely interesting things that can be measured in  DVCS experiments.  Perhaps the most striking is the Ji sum rule[39] which gives the quark contribution, spin plus orbital, to the spin of the proton.  However, the Ji sum rule is extremely hard to measure.  What has come up more recently are some very nice ideas on using DVCS to determine the spatial distribution, in transverse directions, of the quarks in the proton[41-45].  These experiments are not so difficult to do and they would furnish genuinely new information on the structure of the proton.  The idea is simple.

Let $A(x, \Delta_\perp^2, Q^2)$ be the amplitude illustrated in Fig.7 with $x = Q^2/s.$  Then the elastic cross-section is

\begin{equation}
{d\sigma\over d\Delta_\perp^2} \propto \vert A \vert^2.
\end{equation}

\noindent In the approximation that \ A\ is purely real (One can also take into account the imaginary part of \ A\ in a more careful treatment.)

\begin{equation}
A(x, \Delta_\perp^2, Q^2) \propto {\sqrt{{d\sigma\over d\Delta_\perp^2}}}.
\end{equation}

\noindent One can go from momentum transfer to a transverse coordinate basis (impact parameter basis) by

\begin{equation}
A(x, b_\perp, Q^2) \propto \int  d^2\Delta e^{-ib_\perp\cdot\Delta_\perp}{\sqrt{{d\sigma\over d\Delta_\perp^2}}},
\end{equation}

\noindent and $A(x, b_\perp, Q^2)$ gives the impact parameter dependence of the quarks in the light-cone wavefunction.

\section{Golec-Biernat W\"usthoff models and beyond}

The Color Glass Condensate-Saturation idea is what drives much of the theoretical interest in small-x physics.  However, saturation is not so easy to see directly, mainly for two reasons.  (i) While the $F_2$ structure function can be viewed as the scattering of a high-energy dipole on a  proton, the size of the dipole is not fixed, see (10), in terms of the kinematics of the reaction.  (The dipole size is not in general on the order of $1/Q$ in $F_2$ although that is the case for $F_L$ which has yet to be well measured.)  (ii)  The $F_2$ structure function involves an average over impact parameters.  Thus there may be saturation at the center of the proton, but the averaging gives much weight to the edges of the proton where saturation is not present.

The Golec-Biernat W\"usthoff model[9] has played an important role in trying to understand to what extent saturation effects are present in the present HERA data.  The Golec-Biernat W\"usthoff model uses a formula much like (10)

\begin{equation}
F_2={Q^2\over 8\pi^2\alpha_{em}} \int d^2x_\perp \int_0^1dz \sum_f e_f^2\vert\psi_T(x_\perp, z, Q)\vert^2 \sigma_02(1-S)
\end{equation}

\noindent except that the integral over impact parameters is replaced by $\sigma_0$ and \ S\ is paramaterized as

\begin{equation}
S = e^{-x_\perp^2\bar{Q}_s^2/4}
\end{equation}

\noindent with

\begin{equation}
1/R_0^2 = \bar{Q}_s^2 = 1 GeV^2\left(x_0/x\right)^\lambda .
\end{equation}

\noindent $\psi_T$ is the lowest order quark-antiquark wavefunction of the virtual photon.  The model has three parameters

\begin{equation}
\sigma_0 = 23 mb,\ \ \ \ \ \ \ \ \  \lambda = 0.3\ \ \ \ \ {\rm and}\ \ \ \ x_0=5x10^{-4}.
\end{equation}

The Golec-Biernat and W\"usthoff model gives good fits to small-x and moderate $Q^2$ data for $F_2$ and for the diffractive structure function $F_2^D.$  It also suggests a scaling law[46]

\begin{equation}
F_2=F_2(Q^2/\bar{Q}_s^2(x))
\end{equation}

\noindent which seems to work quite well over a surprisingly wide range of \ $x$\  and $Q^2.$

Nevertheless, the model has serious shortcomings.  Impact parameters are still averaged over.  The Gaussian form for  \ $S$\ is not well-motivated.  There are no scaling violations so that (28) does not merge smoothly with forms required by the DGLAP equation at a larger values of $Q^2.$  There have been a number of modifications of the G-B W model which aim at correcting these shortcomings[47-50].  I will briefly discuss only two of them.

Bartels, Golec-Biernat and Kowalski[47] keep the basic structure of the G-B and  W  model (28), but take\   $S$\ in terms of an eikonalized dipole-proton scattering

\begin{equation}
S = exp\{-{1\over 2}\sigma_{dp}/R^2\}
\end{equation}

\noindent where  \ $R$\ is a fixed radius and $\sigma_{dp}$ is given by

\begin{equation}
\sigma_{dp} = {\pi^2\alpha\over N_c} x_\perp^2 x G(x, \mu^2)
\end{equation}

\noindent with

\begin{equation}
\mu^2 = \mu_0^2 + 4/x_\perp^2
\end{equation}

\noindent where $\mu_0$ is a fixed mass. Using (33) in (28) gives a smooth matching onto DGLAP evolution, at least in the leading order DGLAP formalism, and significantly improves fits to the data in the larger part of the $Q^2$ range where the model might apply.

Kowalski and Teaney[50] go one step further.  They keep the form (34) and in addition add an impact parameter variable to the fits.  Thus they go back to a formula identical to (10), and a corresponding formula for diffractive vector meson production, but where now

\begin{equation}
S(x_\perp, b_\perp, x) = exp\{-{1\over 2}\sigma_{dp} T(b_\perp)\}
\end{equation}
 
\noindent with $T(b_\perp)$ given in terms of a  convolution of a Gaussian and a Yukawa form. The saturation momentum now is defined by

\begin{equation}
S^2(x_\perp^2=2/\bar{Q}_s^2, b_\perp,x) = 1/e
\end{equation}

\noindent where, of course the $1/e$ on the right-hand side of (37) is somewhat arbitrary but chosen so that (29) agrees with (36).

This model gives good fits to $F_2$ data and to data for $J/\psi$ diffractive production.  Kowalski and Teaney find that $T(b)$ is much more sharply peaked than electric charge distributions in the proton. Most of the small-x gluons are located in a radius of $b_\perp\simeq 0.6 fm.$  Also the saturation momentun, the quark saturation momentum $\bar{Q}_s,$ comes out to be considerably smaller that that found in the Golec-Biernat and W\"usthoff analysis and in better agreement what seems to be emerging from RHIC phenomenology.

\section{Some miscellaneous topics}

\subsection{Next-to-next-to-leading order corrections[51]}

Major efforts are being made to calculate NNL corrections for many processes.  This means three-loop corrections for anomalous dimensions and two-loop corrections otherwise. This should bring the accuracy of QCD predictions for hard processes to on the order of 5\% and in some  cases even better.  This is an important accomplishment and shows that in many circumstances QCD has become a very precise theory.

\subsection{Non-global observables[52,53]}

The Kovchegov equation arose as an approximate equation for dealing with high-energy scattering when unitarity corrections are important.  Now, remarkably, the same equation seems to appear in a very different context[54,55], that of properties of jet decays.  In particular consider two (nearly) back-to-back jets arising from, say, $e^+e^-$ annihilation. Call $\Sigma_{ab}$ the fraction of decays in which an amount of energy less than $E_{out}$ is deposited outside a cone of size  $\theta$ about the jets $a$ and $b$.  The situation is illustrated in Fig.8.  This is a non-global observable because one requires less than a certain amount of energy be deposited in a  specific region which here has been taken to be outside the cone $\theta .$  At very high center of mass energy,$E$, of the two jets it is clear that $\Sigma$ will be very small while $\Sigma$ should be near $1$ when $E$  is not much bigger than $E_{out}$, at least so long as $\theta$ is not too small.

\begin{figure}
\begin{center}
\centerline{\epsfbox[0 0 124 56]{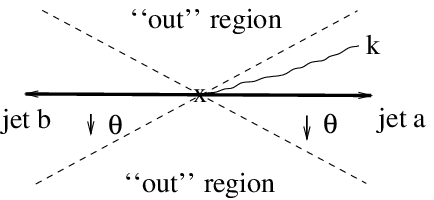}}
\caption{}
\end{center}
\end{figure}

In the large $N_c$ limit one can derive an equation for $\Sigma$ in a very simple way[54,55].  Imagine that $E_{out}$ and  $\theta$ are fixed.  Call $Y=\ell n E/E_{out}$.  Suppose one increases  $Y$ by an amount $dY$.  What can happen is that a gluon at high-energy now has the possibility of being emitted which by energy conservation could not happen previously.  This gluon, $k,$ in Fig.8 can be a real gluon inside the cone region, as illustrated in the figure or it can be a virtual correction.  The real emission means that softer, and later in time, emissions now come not from the dipole $(a,b)$ but either from an $(a,k)$ dipole or from a (k,b) dipole, and in the large $N_c$ limit emissions from these two dipoles do not intefere.  Thus, putting in the details of the gluon emission and calling $\Delta = {\alpha N_c\over \pi} Y,$[54,55],

\begin{equation}{d\Sigma_{ab}(\Delta)\over dY}= - r_{ab} \Sigma_{ab} (\Delta) + \int_{in}{d\Omega_k\over 4\pi} w_{ab}(k)\cdot \left[\Sigma_{ak}(\Delta) \Sigma_{kb}(\Delta) - \Sigma_{ab}(\Delta)\right].
\end{equation}

\noindent The integral on the right-hand side of (38) is restricted to the region inside the $\theta$-cone about the jets  $a$ and $b.$  The first term on the right-hand side of (38) comes about because of the restriction of the virtual term to be inside the cone region.  Thus

\begin{equation} 
r_{ab} = \int_{out}{d\Omega_k\over 4\pi} w_{ab}(k)
\end{equation}
 
\noindent where the integral in (39) is the region outside the cones.  Finally

\begin{equation}
w_{ab}(k) = {1-cos \theta_{ab}\over (1-cos \theta_{ak})(1-cos \theta_{kb})} \simeq {2 \theta_{ab}^2\over \theta_{ak}^2\theta_{kb}^2}
\end{equation}

\noindent where the term on the far right-hand side of (40) is the small angle approximation, which approximation will always dominate the large $\Delta$ behavior of $\Sigma.$  In the small angle approximation, where the jets  $a$  and $b$  are close together and in one of the two original cones Eq.(38) becomes the Kovchegov equation.  Whether this is an accident or whether there is a profound, and unexpected, relationship between certain properties of jet decays and high-energy scattering remains to be understood.

\end{document}